\documentclass[conference]{ieeetran}
\IEEEoverridecommandlockouts
\usepackage{amsmath,amssymb,amsfonts}
\usepackage{amsthm}
\usepackage{bm}
\usepackage{graphicx}
\usepackage{textcomp}
\usepackage{mathptmx}
\usepackage{amsmath}
\usepackage{tabularx}
\usepackage{subfigure}
\usepackage{xcolor}
\usepackage{cite}
\usepackage{multirow}

\graphicspath{{./images/}}
    
\newtheorem{assumption}{Assumption}
\newtheorem{theorem}{Theorem}

\newtheorem{proposition}{Proposition}

\newtheorem{definition}{Definition}

\usepackage{url} 
\usepackage{algorithm,algpseudocode}

\def\BibTeX{{\rm B\kern-.05em{\sc i\kern-.025em b}\kern-.08em
    T\kern-.1667em\lower.7ex\hbox{E}\kern-.125emX}}

\begin{document}

\title{Resilient and Distributed Discrete Optimal Transport with Deceptive Adversary: A Game-Theoretic Approach}

\author{Jason Hughes and Juntao Chen
\thanks{The authors are with the Department of Computer and Information Sciences, Fordham University, New York, NY, 10023 USA. E-mail: \{jhughes50,jchen504\}@fordham.edu}
\thanks{This research was supported in part by a Faculty Research Grant from Fordham Office of Research.}}


\maketitle

\begin{abstract}
Optimal transport (OT) is a framework that can be used to guide the optimal allocation of a limited amount of resources. The classical OT paradigm does not consider malicious attacks in its formulation and thus the designed transport plan lacks resiliency to an adversary. To address this concern, we establish an OT framework that explicitly accounts for the adversarial and stealthy manipulation of participating nodes in the network during the transport strategy design. Specifically, we propose a game-theoretic approach to capture the strategic interactions between the transport planner and the deceptive attacker. We analyze the properties of the established two-person zero-sum game thoroughly. We further develop a fully distributed algorithm to compute the optimal resilient transport strategies, and show the convergence of the algorithm to a saddle-point equilibrium. Finally, we demonstrate the effectiveness of the designed algorithm using case studies.
\end{abstract}


\section{Introduction}

Optimal transport (OT) is a centralized framework that can be leveraged to design efficient resource distribution and matching schemes \cite{galichon2016econ},\cite{bayat2016matching}. The OT framework captures heterogeneous constraints between the resource suppliers and receivers  and it has been used in various applications, such as the distribution of raw materials to manufacturers, dispatching of power restoration facilities in disaster affected neighborhoods, and matching between employees and tasks in an organization. 

Under the standard OT paradigm, the planner designs the resource allocation scheme that maximizes the aggregated utility of all participants \cite{zhang2019consensus,jhughes2021fair}. The classical framework does not consider that the resource suppliers and receivers could be compromised by an attacker whose goal is to disrupt the resource allocation efficiency. To this end, our goal is to develop a more robust transport strategy by using a game-theoretic framework \cite{basar1998dynamic} that captures the interactions between the transport planner and the adversary. Specifically, the planner designs the transport plan that maximizes the social utility by anticipating the compromise of a set of participating nodes by the adversary. In comparison, the attacker's objective is to minimize the aggregated utility of all the nodes under the transport plan. The attacker is stealthy as it will not modify the node's preference information in an arbitrary manner but considers threshold and magnitude constraints during decision-making. The considered scenario is related to the resilient resource allocation under adversarial attacks in literature, including jamming attack\cite{garnaev2014fair}, network topology attack \cite{shao2020distributed}, and data falsification attack \cite{chen2016joint}.

The transport network that the resources are distributed over becomes more complex with a growing number of participants (e.g., resource suppliers and receivers), which can be observed from real-world applications. This large-scale feature of the OT problem gives rise to another concern on the centralized computation of the optimal transport scheme. The required computation for centralized planning grows exponentially with the number of participants in the framework. Thus, our goal is to develop a distributed algorithm for resilient resource transport such that the centralized planner is not necessary. We leverage alternating direction method of multipliers (ADMM) technique \cite{boyd2011distributed} to achieve the distributed transport strategy design. One feature of the designed ADMM-based distributed algorithm is that each participant only needs to solve its own problem and exchange the results with the corresponding connected agents, which enables parallel updates on the transport solution.

To be resilient to strategic attacks, we develop a best response type of algorithm that accounts for the adversarial compromise on the node's preference data. We focus our attention on the scenarios when a set of targets (i.e., resource receivers) are compromised. Thus, in the algorithm, each deceptive target determines its resource requests from the connected source nodes and its manipulations on the preference data. During the iterative update, each target in the network proposes either a truthful solution or an adversarial solution depending on whether the target node is attacked. Comparatively, the source nodes with the goal of maximizing their utility do not respond to the attacks directly but in an implicit manner when computing the transport strategy. This feature can be observed in the designed distributed resilient algorithm. Specifically, at each round of the updates, every pair of source and target nodes propose a resource allocation scheme that is closer to the average of their previous solutions. It indicates that, as the negotiation process proceeds, the sources inherently consider the adversarial impacts by the attacked nodes by this average term to reach a consensus.

The contributions of this paper are summarized as follows. \begin{enumerate}
    \item We establish an adversarial discrete optimal transport framework using a game-theoretic approach that captures the strategic interactions between the resource planner and the attacker. 
    \item We develop an ADMM-based distributed algorithm for computing the optimal transport strategies in the adversarial environment, where the obtained strategy is resilient to the deceptive attacks. 
    \item We show the convergence of the proposed distributed algorithm to a saddle-point equilibrium solution of the established game. We also corroborate the algorithm extensively and show that the algorithm is applicable to large-scale networks due to its distributed nature.
\end{enumerate}

The rest of the paper is organized as follows. Section \ref{sec:problem} formulates a general adversarial OT framework for resource matching. Section \ref{sec:LU} presents a class of adversarial OT problem with linear utilities. Section \ref{sec:analysis} develops a distributed algorithm to compute the resilient optimal transport strategy. Section \ref{sec:case} corroborates the results with case studies, and Section \ref{sec:conclusion} concludes the paper.

\section{Problem Formulation}\label{sec:problem}
In this section, we first present a framework of discrete optimal transport over a network and then formulate an optimal transport problem with adversaries.
\subsection{Discrete Optimal Transport over Network}
We denote by $\mathcal{X}:=\{1, ..., |\mathcal{X}|\}$ the set of destinations/targets that receive the resources, and $\mathcal{Y}:=\{1, ..., |\mathcal{Y}|\}$ the set of origins/sources that distribute resources to the targets in a network. Each source node $y\in\mathcal{Y}$ is connected to a number of target nodes denoted by $\mathcal{X}_y$, representing that $y$ can choose to allocate its resources to a specific group of destinations $\mathcal{X}_y$. Similarly, each target node $x\in\mathcal{X}$ can receive resources from multiple source nodes, and this set of resource suppliers to target $x$ is denoted by $\mathcal{Y}_x$. Note that $\mathcal{X}_y$, $\forall y$ and $\mathcal{Y}_x$, $\forall x$ are nonempty. Otherwise, the corresponding nodes are isolated in the network and do not participant in the resource matching. It can be seen that the resources are transported over a bipartite network, where one side of the network consists of all source nodes and the other includes all destination nodes. This bipartite network is not necessarily complete because of constrained matching policies between participants. An incomplete bipartite graph also models the infeasible transport of resources between certain pairs of source and destination nodes incurred by long transport distance. For convenience, we denote by $\mathcal{E}$ the set including all feasible transport paths in the network, i.e., $\mathcal{E}:=\{\{x,y
\}|x\in\mathcal{X}_y,y\in\mathcal{Y}\}$. Here, $\mathcal{E}$ also refers to the set of all edges in the established bipartite graph for resource transportation.

We next denote by $\pi_{xy}\in\mathbb{R}_+$ the amount of resources transported from the origin node $y\in\mathcal{Y}$ to the destination node $x\in\mathcal{X}$, where $\mathbb{R}_+$ is the set of nonnegative real numbers. Let $\Pi:=\{\pi_{xy}\}_{x\in\mathcal{X}_y,y\in\mathcal{Y}}$ be the designed transport plan. To this end, the centralized optimal transport problem can be formulated as follows:
\begin{equation}\label{OT1:eqn}
\begin{aligned}
    \max_{\Pi}\ \sum_{x\in\mathcal{X}} \sum_{y\in\mathcal{Y}_x}& t_{xy}(\pi_{xy}) + \sum_{y\in\mathcal{Y}} \sum_{x\in\mathcal{X}_y} s_{xy}(\pi_{xy})\\
    \mathrm{s.t.}\quad &\underline{p}_{x}\leq \sum_{y\in\mathcal{Y}_x} \pi_{xy}\leq \bar{p}_{x},\ \forall x\in\mathcal{X},\\
    &\underline{q}_{y}\leq \sum_{x\in\mathcal{X}_y} \pi_{xy}\leq \bar{q}_{y},\ \forall y\in\mathcal{Y},\\
    &\pi_{xy}\geq 0,\ \forall \{x,y\} \in\mathcal{E},
\end{aligned}
\end{equation}
where $t_{xy}:\mathbb{R}_+\rightarrow\mathbb{R}$ and $s_{xy}:\mathbb{R}_+\rightarrow\mathbb{R}$ are utility functions for target node $x$ and source node $y$, respectively. Furthermore, $\bar{p}_x\geq \underline{p}_{x}\geq 0$, $\forall x\in\mathcal{X}$ and $\bar{q}_y\geq \underline{q}_{y}\geq 0$, $\forall y\in\mathcal{Y}$. The constraints $\underline{p}_{x}\leq \sum_{y\in\mathcal{Y}_x} \pi_{xy}\leq \bar{p}_{x}$ and $\underline{q}_{y}\leq \sum_{x\in\mathcal{X}_y} \pi_{xy}\leq \bar{q}_{y}$ capture the limitations on the amount of requested and transferred resources at the target $x$ and source $y$, respectively. 

We have the following assumption on the utility functions.
\begin{assumption}\label{assump:1}
The utility functions $t_{xy}$ and $s_{xy}$ are concave and monotonically increasing on $\pi_{xy}$, $\forall x\in\mathcal{X},\forall y\in\mathcal{Y}$.
\end{assumption}

Recall that a function $f$ is concave on an interval if for any $x$ and $y$ in the interval and for any $\theta\in[0,1]$, $f((1-\theta)x+\theta y)\geq (1-\theta)f(x)+\theta f(y)$. A rich class of functions satisfy the conditions in Assumption \ref{assump:1}. For example, the utility functions $t_{xy}$ and $s_{xy}$ can be linear on $\pi_{xy}$, indicating a linear growth of benefits on the amount of transferred and consumed resources. These two functions can also admit a logarithmic form, capturing that the marginal utility decreases as the amount of transported resources increase.

\subsection{Adversarial Optimal Transport}
The attacker's goal is to minimize the aggregated transport utility by compromising the preference coefficients in the target's utility functions (which can happen at the information exchange stage). Specifically, the parameters in the utility function $t_{xy}$ are compromised, for $x\in\mathcal{X}_a$, $y\in\mathcal{Y}_x$, where $\mathcal{X}_a$ denotes a subset of adversarial receiver nodes. Then, $\mathcal{X}_o:=\mathcal{X}\setminus\mathcal{X}_a$ is the set of uncompromised targets. We denote by $\tilde{t}_{xy,\xi_{xy}}$ the modified utility under the attack, where $\xi_{xy}$ represents the magnitude of the adversarial modifications on the corresponding parameters. For example, when the utility function admits a linear form as $t_{xy}(\pi_{xy}) = \delta_{xy}\pi_{xy}$, where $\delta_{xy} > 0$ is a parameter, the compromised utility form under the deception attack becomes $\tilde{t}_{xy,\xi_{xy}}(\pi_{xy}) = (\delta_{xy}+\xi_{xy})\pi_{xy}$. As another example, when $t_{xy}$ takes a form of $t_{xy}(\pi_{xy}) = \delta_{xy}\min(\zeta_x,\pi_{xy})$, where $\zeta_x$ denotes a threshold after which the benefit of consuming more resources for target $x$ does not increase, the compromised utility form can be constructed as $\tilde{t}_{xy,\xi_{xy}}(\pi_{xy}) = (\delta_{xy}+\xi_{xy,1})\min\{\zeta_x+\xi_{xy,2},\pi_{xy}\}$. As another example, when $t_{xy}$ takes a form of $t_{xy}(\pi_{xy}) = \delta_{xy}\min(\zeta_{xy},\pi_{xy})$, where $\zeta_{xy}$ denotes a threshold after which the benefit of consuming more resources for target $x$ from source $y$ does not increase, the compromised utility form can be constructed as $\tilde{t}_{xy,\xi_{xy}}(\pi_{xy}) = (\delta_{xy}+\xi_{xy,1})\min\{\zeta_{xy}+\xi_{xy,2},\pi_{xy}\}$. In this scenario, the attacker's action includes both $\xi_{xy,1}$ and $\xi_{xy,2}$, $\forall x\in\mathcal{X}_a$, $y\in\mathcal{Y}_x$. For a general scenario, we denote by  $\Xi:=\{\xi_{xy}\}_{x\in\mathcal{X}_a,y\in\mathcal{Y}_x}$ the attacker's deceptive strategy. Then, the adversarial optimal transport can be formulated as follows.
\begin{equation}\label{OT2_deception:eqn}
\begin{aligned}
    \max_{\Pi}\min_{\Xi}\ &\sum_{x\in\mathcal{X}_a} \sum_{y\in\mathcal{Y}_x} t_{xy}(\pi_{xy}) + \sum_{y\in\mathcal{Y}} \sum_{x\in\mathcal{X}_y} s_{xy}(\pi_{xy})\\
    &+\sum_{x\in\mathcal{X}_o} \sum_{y\in\mathcal{Y}_x} \tilde{t}_{xy,\xi_{xy}}(\pi_{xy})+\sum_{x\in\mathcal{X}_a}\sum_{y\in\mathcal{Y}_x}l(\xi_{xy})\\
    \mathrm{s.t.}\quad &\underline{p}_{x}\leq \sum_{y\in\mathcal{Y}_x} \pi_{xy}\leq \bar{p}_{x},\ \forall x\in\mathcal{X},\\
    &\underline{q}_{y}\leq \sum_{x\in\mathcal{X}_y} \pi_{xy}\leq \bar{q}_{y},\ \forall y\in\mathcal{Y},\\
    &\pi_{xy}\geq 0,\ \forall \{x,y\} \in\mathcal{E},\\
    & \bm{\xi}_{x}\in\mathcal{A}_x,\ \forall x\in\mathcal{X}_a,
\end{aligned}
\end{equation}
where $\bm{\xi}_{x}:=[\xi_{x1},\xi_{x2},...,\xi_{x|\mathcal{Y}_x|}]$, for $x\in\mathcal{X}_a$; and $\mathcal{A}_x$ is the attacker's feasible action set on the target node $x\in\mathcal{X}_a$.and $l:\mathbb{R}\rightarrow \mathbb{R}_+$ is a function capturing the cost of the attack.

\textit{Remark:} The solution to the adversarial OT problem is related to the robust OT design. Robust OT also admits a minimax formulation but its goal is to find an optimal solution in the presence of structural and known uncertainties. Comparatively, in the adversarial OT, such uncertainty is replaced by strategic attacks, and the designed transport plan should be resistant to adversarial manipulations.

\section{Adversarial Optimal Transport under Linear Utilities}\label{sec:LU}

In this section, we consider utility functions admitting a linear form for both the sender and receiver. Specifically, $t_{xy}(\pi_{xy}) = \delta_{xy}\pi_{xy}$ and $s_{xy}(\pi_{xy}) = \gamma_{xy}\pi_{xy}$, where $\delta_{xy},\gamma_{xy}\in\mathbb{R}_+$. To design the optimal transport plan, the transport planner needs to know the utility parameters including $\delta_{xy}$, $\gamma_{xy}$, $\forall x\in\mathcal{X}, y\in\mathcal{Y}_x$. Thus, the source nodes and target nodes need to report their parameters, and one way to achieve this is through communications. The wireless channel enabling the communication is vulnerable to cyber attacks. The attacker can disrupt the communication by various techniques, such as jamming and distributed denial of service attacks. Therefore, it is imperative for the central planner to develop resilient transport strategies under the adversarial environment. In the considered scenario, we assume that the attacker is capable to compromise a subset of receiver nodes in the network, denoted by $\mathcal{X}_a$. One interpretation is the nodes in $\mathcal{X}_a$ do not have a secure communication protocol with the central planner. In comparison, the nodes in the set $\mathcal{X}_o=\mathcal{X}\setminus\mathcal{X}_a$ are able to set up high-confidence communication channels and hence are secure from adversarial attacks.

The attacker compromises the sensitive data $\delta_{xy}$, $x\in\mathcal{X}_a,y\in\mathcal{Y}_x$, reported by the vulnerable target nodes and stealthily modify them to new values aiming to decrease the social utility of resource transportation. The adversarial disruption can be regarded as a data poisoning attack, under which the data point $\delta_{xy}$ is changed to $\tilde{\delta}_{xy} := \delta_{xy} + \xi_{xy}$. for $x\in\mathcal{X}_a,y\in\mathcal{Y}_x$. Here, $\xi_{xy}$ denotes the action of the attacker, representing the magnitude of data modification to the particular data point $\delta_{xy}$. For convenience, we follow the notations in \eqref{OT2_deception:eqn}, where $\Xi$ denotes the attacker's malicious manipulations on the data points and $\bm{\xi}_{x}$ is the attackers action on the target node $x\in\mathcal{X}_a$.

To this end, the adversarial OT can be formulated in the following max-min format:
\begin{equation}\label{OTA:eqn}
\begin{aligned}
    \max_{\Pi}\min_{\Xi}\ &U(\Pi,\Xi)=\sum_{x\in\mathcal{X}_o} \sum_{y\in\mathcal{Y}_x} \delta_{xy}\pi_{xy} + \sum_{y\in\mathcal{Y}} \sum_{x\in\mathcal{X}_y} \gamma_{xy}\pi_{xy}\\
    &+\sum_{x\in\mathcal{X}_a} \sum_{y\in\mathcal{Y}_x} (\delta_{xy}+\xi_{xy})\pi_{xy}+c_a\sum_{x\in\mathcal{X}_a} \Vert \bm{\xi}_{x}\Vert_1\\
    \mathrm{s.t.}\quad &\underline{p}_{x}\leq \sum_{y\in\mathcal{Y}_x} \pi_{xy}\leq \bar{p}_{x},\ \forall x\in\mathcal{X},\\
    &\underline{q}_{y}\leq \sum_{x\in\mathcal{X}_y} \pi_{xy}\leq \bar{q}_{y},\ \forall y\in\mathcal{Y},\\
    &\pi_{xy}\geq 0,\ \forall \{x,y\} \in\mathcal{E},\\
    & \bm{\xi}_{x}\in\mathcal{A}_x,\ \forall x\in\mathcal{X}_a,
\end{aligned}
\end{equation}
where $c_a\in\mathbb{R}_+$ is a non-negative cost coefficient and $\mathcal{A}_x$ is the feasible action set of the attacker on target node $x$, $x\in\mathcal{X}_a$. $U$ is the objective value under strategies $\Pi$ and $\Xi$. The term $c_a\sum_{x\in\mathcal{X}_a} \Vert \bm{\xi}_{x}\Vert_1$ captures the cost of the attack. The sparsity induced by the $l_1$ norm is a convex approximation of the $l_0$ norm \cite[Chapter 6]{boyd2011distributed} and indicates that the attacker has constraints on the number of compromise of utility parameters at a particular node $x\in\mathcal{X}_a$. The attacker is a minimizer of \eqref{OTA:eqn} as its goal is to minimize the aggregated transport utility reflected by the first three terms in the objective function $U$ while using the least costly attack scheme captured by the last term in $U$.

If the attacker modifies all the data parameters significantly, it is easy for the planner to detect such adversarial perturbations. Also, the data $\tilde{\delta}_{xy}$ after compromise should still be non-negative. Otherwise, the deception can be identified straightforwardly. Thus, the action set $\mathcal{A}_x$ needs to be carefully modeled to capture the attacker's deceptive behavior. One form of $\mathcal{A}_x$ can be chosen as follows:
\begin{equation}
    \mathcal{A}_x = \{\bm{\xi}_x\vert \Vert\bm{\xi}_x \Vert_2^2\leq \kappa_x, \bm{\xi}_x+\bm{\delta}_x\geq \bm{0}\},\ x\in\mathcal{X}_a,
\end{equation}
where $\kappa_x\in\mathbb{R}_+$ denotes the upper limit of the standard norm of adversarial modifications at the target node $x\in\mathcal{X}_a$ by the attacker; $\bm{\delta}_x:=[\delta_{x1};\delta_{x2};...;\delta_{x|\mathcal{Y}_x|}]$; and $\bm{0}$ is a zero vector with appropriate dimension.

Problem \eqref{OTA:eqn} can be seen as a two-person zero-sum game denoted by $G$, where the transport planner is a maximizer and the attacker is a minimizer. The solution to the game $G$ is characterized by Nash equilibrium which predicts the outcome of the optimal transport strategy under adversarial environment. The formal definition of the Nash equilibrium strategy \cite{basar1998dynamic} is presented as follows.

\begin{definition}[Nash Equilibrium]
The strategy pair $\{\Pi^*,\Xi^*\}$ is a saddle-point Nash equilibrium of game $G$ if  
\begin{equation}
    U(\Pi,\Xi^*)\leq U(\Pi^*,\Xi^*)\leq U(\Pi^*,\Xi), \ \forall\ \Pi, \Xi
\end{equation}
where $U$ is the objective function in \eqref{OTA:eqn}.
\end{definition}

Solving game $G$ requires to address the formulated max-min problem \eqref{OTA:eqn}. Specifically, both the central planner and the attacker need to compute their solutions holistically. This centralized computation paradigm does not scale well as the number of nodes in the transport network becomes enormous. Furthermore, to compute the solution $\Pi$, the central planner is required to have a complete information on the transport network, including the sensitive parameters of all participants' preferences. Thus, it is imperative to design a computationally efficient mechanism to solve game $G$. Our subsequent goal is to develop a distributed algorithm to compute the equilibrium transport strategy which also preserves the privacy of the participants to some extent.

\section{Analysis and Distributed Algorithm}\label{sec:analysis}
In this section, we aim to design a holistic and fully distributed algorithm to compute the optimal strategies of the attacker and the participants in the transport network.

\subsection{Equivalence between Max-Min and Minimax Problems}
Before designing the algorithm, we prove that the formulated max-min problem \eqref{OTA:eqn} is equivalent to its minimax counterpart and hence show the existence of Nash equilibrium to game $G$. Specifically, we have the following results.

\begin{proposition}\label{prop:minimax_equi}
The max-min problem \eqref{OTA:eqn} yields the same solution as its minimax counterpart, i.e., $\min_{\Xi}\max_{\Pi}\ U(\Pi,\Xi)$ subject to the same set of the constraints as in \eqref{OTA:eqn}. Thus, there exists saddle point Nash equilibrium to game $G$. However, such equilibrium is not necessarily unique.
\end{proposition}
\begin{proof}
The equivalence between max-min and minimax problems directly follows from the von Neumann's minimax theorem \cite{nikaido1954neumann}. As the objective function $U$ is not strictly concave in $\Pi$ and not strictly convex in $\Xi$, the Nash equilibrium is not necessarily unique \cite[Chapter 4]{basar1998dynamic}.
\end{proof}


Note that Proposition \ref{prop:minimax_equi} facilitates a convenient design of efficient mechanisms called best-response dynamics in finding the equilibrium strategies. We will describe this approach in detail in the ensuing sections. 

\subsection{Distributed Updates on the Deception Strategy}
The attacker deceives the transport planner by compromising $\delta_{xy}$, $x\in\mathcal{X}_a, y\in\mathcal{Y}_x$, strategically. As the attacker's goal is to minimize $U$, a smaller $\tilde{\delta}_{xy}$ (hence a smaller $\delta_{xy}$) will decrease the utility at the corresponding target node as indicated by the term $\sum_{x\in\mathcal{X}_a} \sum_{y\in\mathcal{Y}_x} (\delta_{xy}+\xi_{xy})\pi_{xy}$. However, simply modifying the values of all $\delta_{xy}$, $\forall x\in\mathcal{X}_a, y\in\mathcal{Y}_x$, to their minimum does not guarantee to minimize $U$. One reason is that the transport strategy will be changed under the attack. Though the value of term $\sum_{x\in\mathcal{X}_a} \sum_{y\in\mathcal{Y}_x} (\delta_{xy}+\xi_{xy})\pi_{xy}$ decreases, other terms such as $\sum_{x\in\mathcal{X}_o} \sum_{y\in\mathcal{Y}_x} \delta_{xy}\pi_{xy}$ and $ \sum_{y\in\mathcal{Y}} \sum_{x\in\mathcal{X}_y} \gamma_{xy}\pi_{xy}$ may increase under the attack. Thus, the attacker's deceptive strategy is nontrivial to devise.

In the following, we describe how to leverage best-response dynamics to compute the strategy. Specifically, the attacker updates its decision $\Xi$ by fixing the transport planner's strategy $\Pi'=\{\pi_{xy}'\}_{x\in\mathcal{X}_y,y\in\mathcal{Y}}$. In this regard, the first two terms in the objective function $U(\Pi,\Xi)$ and the first three constraints in \eqref{OTA:eqn} can be safely ignored as they are irrelevant with the attacker's deceptive strategy design. Thus, the attacker solves the following optimization program:
\begin{equation}\label{OTA_attacker:eqn}
\begin{aligned}
    \min_{\Xi}\ & \sum_{x\in\mathcal{X}_a} \sum_{y\in\mathcal{Y}_x} \xi_{xy}\pi_{xy}' +c_a\sum_{x\in\mathcal{X}_a} \Vert \bm{\xi}_{x}\Vert_1\\
    \mathrm{s.t.}\quad
    & \bm{\xi}_{x}\in\mathcal{A}_x,\ \forall x\in\mathcal{X}_a.
\end{aligned}
\end{equation}
The attacker can design the optimal deceptive strategy $\Xi^*$ in a distributed fashion. First, we observe that the cost function in \eqref{OTA_attacker:eqn} is decoupled across vulnerable target nodes. Then, the optimal $\bm{\xi}_x^*$, $\forall x\in\mathcal{X}_a$, can be obtained separately. Solving \eqref{OTA_attacker:eqn} is thus equivalent to addressing $|\mathcal{X}_a|$ sub-problems as follows, for $x\in\mathcal{X}_a$,
\begin{equation}\label{OTA_attacker_subp:eqn}
\begin{aligned}
    \min_{\bm{\xi}_x}\ &  \sum_{y\in\mathcal{Y}_x} \xi_{xy}\pi_{xy}'+c_a \Vert \bm{\xi}_{x}\Vert_1\\
    \mathrm{s.t.}\quad
    & \bm{\xi}_{x}\in\mathcal{A}_x.
\end{aligned}
\end{equation}We can further rewrite \eqref{OTA_attacker_subp:eqn} in the following form, for $x\in\mathcal{X}_a$:
\begin{equation}\label{OTA_attacker_subp_2:eqn}
\begin{aligned}
    \min_{\bm{\xi}_x,\bm{\chi}_x}\ &  \sum_{y\in\mathcal{Y}_x} \xi_{xy}\pi_{xy}'+ \bm{1}^{\mathsf{T}} \bm{\chi}_x\\
    \mathrm{s.t.}\quad
    & \bm{\xi}_{x}\in\mathcal{A}_x,\\
    & c_a\bm{\xi}_{x}\leq \bm{\chi}_x,\\
    & c_a\bm{\xi}_{x}\geq -\bm{\chi}_x,
\end{aligned}
\end{equation}
where $\bm{1}$ is a vector of appropriate dimension with all ones; $\mathsf{T}$ denotes the transpose operator; and $\bm{\chi}_x$ is an auxiliary $|\mathcal{Y}_x|$-dimensional decision variable. Note that the objective function in $\eqref{OTA_attacker_subp_2:eqn}$ is linear and the constraints are convex, and thus $\eqref{OTA_attacker_subp_2:eqn}$ can be solved efficiently.

\textit{Equivalence between problems \eqref{OTA_attacker_subp:eqn} and \eqref{OTA_attacker_subp_2:eqn}}: First, we can rewrite $c_a \Vert \bm{\xi}_{x}\Vert_1$ as $\sum_{i=1}^{|\bm{\xi}_{x}|}\mathrm{abs}(c_a \bm{\xi}_{x,i})$, where $\bm{\xi}_{x,i}$ is the $i$-th element of $\bm{\xi}_{x}$ and $\mathrm{abs}(\cdot)$ denotes an operator of taking the absolute value. Thus, the objective function of (7) can be recast as $\sum_{y\in\mathcal{Y}_x} \xi_{xy}\pi_{xy}'+ \sum_{i=1}^{|\bm{\xi}_{x}|}\mathrm{abs}(c_a \bm{\xi}_{x,i})$. We then introduce an auxiliary variable $\bm{\chi}_x$ with a same dimension as $\bm{\xi}_x$ that satisfies the condition $\mathrm{abs}(c_a \bm{\xi}_{x,i})\leq \bm{\chi}_{x,i}$, $\forall i$. Then the optimization problem
\begin{equation*}
\begin{aligned}
    \min_{\bm{\xi}_x}\ &  \sum_{y\in\mathcal{Y}_x} \xi_{xy}\pi_{xy}'+\sum_{i=1}^{|\bm{\xi}_{x}|}\mathrm{abs}(c_a \bm{\xi}_{x,i})\\
    \mathrm{s.t.}\quad
    & \bm{\xi}_{x}\in\mathcal{A}_x,
\end{aligned}
\end{equation*}
can be reformulated as 
\begin{equation*}
\begin{aligned}
    \min_{\bm{\xi}_x,\bm{\chi}_x}\ &  \sum_{y\in\mathcal{Y}_x} \xi_{xy}\pi_{xy}'+\sum_{i=1}^{|\bm{\chi}_{x}|}\bm{\chi}_{x,i}\\
    \mathrm{s.t.}\quad
    & \bm{\xi}_{x}\in\mathcal{A}_x,\\
    & \mathrm{abs}(c_a \bm{\xi}_{x,i})\leq \bm{\chi}_{x,i},\ \forall i=1,..., |\bm{\chi}_{x}|.
\end{aligned}
\end{equation*}
Note that $\sum_{i=1}^{|\bm{\chi}_{x}|}\bm{\chi}_{x,i}$ is equivalent to $\bm{1}^{\mathsf{T}}\bm{\chi}_{x}$. In addition, $ \mathrm{abs}(c_a \bm{\xi}_{x,i})\leq \bm{\chi}_{x,i}$ can be written as $-\bm{\chi}_{x,i} \leq c_a \bm{\xi}_{x,i}\leq \bm{\chi}_{x,i}$, $\forall i$. Putting it in a vector form yields $-\bm{\chi}_x \leq c_a\bm{\xi}_{x}\leq \bm{\chi}_x$. Thus, we obtain the formulation of \eqref{OTA_attacker_subp_2:eqn}.

\subsection{Distributed Updates on the Transport Strategy}
Under the best-response mechanism, similarly, the transport planner determines the transport strategy by regarding the deceptive strategy $\Xi'=\{\xi_{xy}'\}_{x\in\mathcal{X}_a,y\in\mathcal{Y}_x}$ as fixed. Thus, the planner can omit the last term in the objective function $U(\Pi,\Xi)$ and the last constraint in \eqref{OTA:eqn} when making the decision. The planner's problem can be formulated as follows. 
\begin{equation}\label{OTA_planner:eqn}
\begin{aligned}
    \max_{\Pi}\ & \sum_{x\in\mathcal{X}_o} \sum_{y\in\mathcal{Y}_x} \delta_{xy}\pi_{xy} + \sum_{y\in\mathcal{Y}} \sum_{x\in\mathcal{X}_y} \gamma_{xy}\pi_{xy}\\
    &+\sum_{x\in\mathcal{X}_a} \sum_{y\in\mathcal{Y}_x} (\delta_{xy}+\xi_{xy}')\pi_{xy}\\
    \mathrm{s.t.}\quad &\underline{p}_{x}\leq \sum_{y\in\mathcal{Y}_x} \pi_{xy}\leq \bar{p}_{x},\ \forall x\in\mathcal{X},\\
    &\underline{q}_{y}\leq \sum_{x\in\mathcal{X}_y} \pi_{xy}\leq \bar{q}_{y},\ \forall y\in\mathcal{Y},\\
    &\pi_{xy}\geq 0,\ \forall \{x,y\} \in\mathcal{E}.
\end{aligned}
\end{equation}

Solving \eqref{OTA_planner:eqn} in a centralized manner requires the transport planner to know all parameters including $\delta_{xy}$ and $\gamma_{xy}$, $\forall \{x,y\}\in\mathcal{E}$. Our next goal is to design a distributed method to compute the optimal $\Pi$ in \eqref{OTA_planner:eqn}.

First, we introduce auxiliary variables $\pi_{xy}^t$ and $\pi_{xy}^s$ denoting the amount of resources requested by target $x$ from source $y$ and source $y$ offering to target $x$, respectively. These two transport plans should be equal to each other to reach a consensus. Thus, we have constraints $\pi_{xy}^t=\pi_{xy}$ and $\pi_{xy}=\pi_{xy}^s$, $\forall\{x,y\}\in\mathcal{E}$. Then, \eqref{OTA_planner:eqn} can be reformulated as follows.
\begin{equation}\label{mainform:eqn}
\begin{aligned}
    \min_{\Pi^t \in \mathcal{F}_t, \Pi^s \in \mathcal{F}_s} -\sum_{x\in\mathcal{X}_o} \sum_{y\in\mathcal{Y}_x} \delta_{xy}\pi_{xy}^t - \sum_{y\in\mathcal{Y}} \sum_{x\in\mathcal{X}_y} \gamma_{xy}\pi_{xy}^s&\\ -\sum_{x\in\mathcal{X}_a} \sum_{y\in\mathcal{Y}_x} (\delta_{xy}+\xi_{xy}')\pi_{xy}^t&\\
    \mathrm{s.t.} \quad \pi_{xy}^t = \pi_{xy}, \forall \{x,y\} \in \mathcal{E},&\\
         \pi_{xy} = \pi_{xy}^s, \forall\{x,y\} \in \mathcal{E},
\end{aligned}
\end{equation}
where $\Pi^t:=\{\pi_{xy}^t\}_{x\in\mathcal{X}_y,y\in\mathcal{Y}}$, $\Pi^s:=\{\pi_{xy}^s\}_{x\in\mathcal{X},y\in\mathcal{Y}_x,}$, $\mathcal{F}_t := \{ \Pi^t | \pi_{xy}^t \geq 0, \underline{p}_x \leq \sum_{y \in \mathcal{Y}_x} \pi_{xy}^t \leq \bar{p}_x,\{x,y\} \in \mathcal{E}\}$, and $\mathcal{F}_s := \{ \Pi^s | \pi_{xy}^s \geq 0, \underline{q}_y \leq \sum_{x \in \mathcal{X}_y} \pi_{xy}^s \leq \bar{q}_y,\{x,y\} \in \mathcal{E}\}$.

From the convex form of the formulation we can obtain the Lagrangian: 
\begin{equation}\label{lagrangian:eqn}
\begin{aligned}
L(\Pi_t,\Pi_s,\Pi,\alpha_{xy}^t,\alpha_{xy}^s) = -\sum_{x\in\mathcal{X}_o} \sum_{y\in\mathcal{Y}_x} \delta_{xy}\pi_{xy}^t - \sum_{y\in\mathcal{Y}} \sum_{x\in\mathcal{X}_y} \gamma_{xy}\pi_{xy}^s \\
-\sum_{x\in\mathcal{X}_a}\sum_{y\in\mathcal{Y}_x} \left( \delta_{xy} + \xi_{xy}' \right) \pi_{xy}^t + \sum_{x\in\mathcal{X}} \sum_{y\in\mathcal{Y}_x} \alpha_{xy}^t \left( \pi_{xy}^t - \pi_{xy} \right) \\ 
+ \sum_{y\in\mathcal{Y}} \sum_{x\in\mathcal{X}_y} \alpha_{xy}^s \left( \pi_{xy} - \pi_{xy}^s \right) + \frac{\eta}{2} \sum_{x\in\mathcal{X}} \sum_{y\in\mathcal{Y}_x} \left( \pi_{xy}^t - \pi_{xy} \right)^2 \\
+\frac{\eta}{2} \sum_{x\in\mathcal{X}} \sum_{y\in\mathcal{Y}_x} \left( \pi_{xy} - \pi_{xy}^s \right)^2.
\end{aligned}
\end{equation}
Here, $\alpha_{xy}^t$ and $\alpha_{xy}^s$ are Lagrangian multipliers associated with the constraints, and $\eta$ is a positive constant. 
\begin{theorem}
We obtain the following steps using the ADMM algorithm to \eqref{mainform:eqn}:
\begin{equation}\label{unsimp_admm1:eqn}
\begin{aligned}
\Pi_{x}^t(k+1) \in \arg \min_{\Pi_{x}^t \in \mathcal{F}_{x}^t} -\sum_{y \in \mathcal{Y}_x} \delta_{xy} \pi_{xy}^t +\sum_{y \in \mathcal{Y}_x} \alpha_{xy}^t (k)\pi_{xy}^t \\+ \frac{\eta}{2} \sum_{y \in \mathcal{Y}_x}\left( \pi_{xy}^t - \pi_{xy}(k) \right)^2,
\end{aligned}
\end{equation}
\begin{equation}\label{unsimp_admm2:eqn}
\begin{aligned}
    \Pi_{x}^t(k+1) \in \arg \min_{\Pi_{x}^t \in \mathcal{F}_{x}^t} -\sum_{y \in \mathcal{Y}_x} \left( \delta_{xy} + \xi_{xy}' \right) \pi_{xy}^t \\ +\sum_{y \in \mathcal{Y}_x} \alpha_{xy}^t (k)\pi_{xy}^t + \frac{\eta}{2} \sum_{y \in \mathcal{Y}_x}\left( \pi_{xy}^t - \pi_{xy}(k) \right)^2,
\end{aligned}
\end{equation}
where we use \eqref{unsimp_admm1:eqn} for $x\in\mathcal{X}_o$ and \eqref{unsimp_admm2:eqn} for $x\in\mathcal{X}_a$.
\begin{equation}\label{unsimp_admm3:eqn}
\begin{aligned}
\Pi_{y}^s(k+1) \in \arg \min_{\Pi_{y}^s \in \mathcal{F}_{y}^s} -\sum_{x \in \mathcal{X}_y} \gamma_{xy} \pi_{xy}^s +\sum_{x \in \mathcal{X}_y} \alpha_{xy}^s(k)\pi_{xy}^s \\ +\frac{\eta}{2} \sum_{x \in \mathcal{X}_y} \left( \pi_{xy}(k) - \pi_{xy}^s \right),
\end{aligned}
\end{equation}
\begin{equation}\label{unsimp_admm4:eqn}
\begin{aligned}
\pi_{xy}(k+1) \in \arg \min_{\pi_{xy}}  \alpha_{xy}^t(k)\pi_{xy} + \alpha_{xy}^s(k)\pi_{xy}\\ + \frac{\eta}{2} (\pi_{xy}^t(k+1) - \pi_{xy})^2 + \frac{\eta}{2} (\pi_{xy} - \pi_{xy}^s(k+1))^2,
\end{aligned}
\end{equation}
\begin{equation}\label{unsimp_admm5:eqn}
\begin{aligned}
\alpha_{xy}^t(k+1) = \alpha_{xy}^t(k) + \eta (\pi_{xy}^t(k+1) - \pi_{xy}(k+1))^2,
\end{aligned}
\end{equation}
\begin{equation}\label{unsimp_admm6:eqn}
\begin{aligned}
\alpha_{xy}^s(k+1) = \alpha_{xy}^s(k) + \eta (\pi_{xy}(k+1) - \pi_{xy}^s(k+1))^2,
\end{aligned}
\end{equation}
where $\Pi^t_{\tilde{x}}=\{\pi_{xy}^t\}_{y\in\mathcal{Y}_x,x=\tilde{x}}$ and $\Pi^s_{\tilde{y}}=\{\pi_{xy}^s\}_{x\in\mathcal{X}_y,y=\tilde{y}}$ denote the transport strategy computed by target node $\tilde{x}$ and source node $\tilde{y}$, respectively. Additionally, we define $\mathcal{F}_{x}^t := \{ \Pi_{x}^t \vert \pi_{xy}^t \geq 0, y \in \mathcal{Y}_x, \underline{p}_x \leq \sum_{y \in \mathcal{Y}_x} \pi_{xy}^t \leq \bar{p}_x\}$ and $\mathcal{F}_{y}^s := \{ \Pi_{y}^s \vert \pi_{xy}^s \geq 0, x \in \mathcal{X}_y, \underline{q}_y  \leq \sum_{x \in \mathcal{X}_y} \pi_{xy}^s \leq \bar{q}_x\}$. 
\end{theorem}
\begin{proof}
Let $\Vec{x} = [\Vec{\Pi}_x^{t\mathsf{T}}, \Vec{\Pi}^\mathsf{T}]^\mathsf{T}$, $\Vec{y} = [\Vec{\Pi}^\mathsf{T}, \Vec{\Pi}_y^{s\mathsf{T}}]^\mathsf{T}$, and $\alpha = [\{{\alpha_{xy}^{s}}\}^\mathsf{T}, \{{\alpha_{xy}^{t}}\}^\mathsf{T}]^\mathsf{T}$, where $\mathsf{T}$ and $\Vec{}$ denote the transpose and vectorization operator. Note that these three vectors are all $2|\mathcal{E}| \times 1$. Now we can write the constraints in \eqref{mainform:eqn} in a matrix form such that $\mathbf{A}\Vec{x} = \vec{y}$, where $\mathbf{A} = [\textbf{I},\textbf{0};\textbf{0},\textbf{I}]$ with $\textbf{I}$ and $\textbf{0}$ denoting the $|\mathcal{E}|$-dimensional identity and zero matrices, respectively. Next, we note that $\Vec{x} \in \mathcal{F}_{\Vec{x}}^t$ and $\Vec{y} \in \mathcal{F}_{\Vec{y}}^s$, where
$
      \mathcal{F}_{\Vec{x}}^t = \{ \Vec{x} | \pi_{xy}^{t} \geq 0, \underline{p}_x \leq \sum_{y \in \mathcal{Y}_x} \pi_{xy}^{t} \leq \bar{p}_x, \{x,y\} \in \mathcal{E} \},\ 
       \mathcal{F}_{\Vec{y}}^s := \{ \Vec{y} | \pi_{xy}^{s} \geq 0, \underline{q}_y \leq \sum_{x \in \mathcal{X}_y} \pi_{xy}^{s} \leq \bar{q}_y, \{x,y\} \in \mathcal{E}  \}.
$
Then, we can solve \eqref{mainform:eqn} using the iterations: 1)
$
    \Vec{x}(k+1) \in \arg \min_{\Vec{x} \in \mathcal{F}_{\Vec{x}}^t} L(\Vec{x},\Vec{y}(k),\alpha(k));
$
2)
$
    \Vec{y}(k+1) \in \arg \min_{\Vec{y} \in \mathcal{F}_{\Vec{y}}^s} L(\Vec{x}(k+1),\Vec{y},\alpha(k));
$
3)
$
    \alpha(k+1) = \alpha(k) + \eta(A\Vec{x}(k+1) - \Vec{y}(k+1)),
$ based on \cite{boyd2011distributed}.
Because we have no couplings among $\Pi_{x}^t, \Pi_{y}^s, \Pi, \alpha_{xy}^{t}$ and $\alpha_{xy}^{s}$, the above iterations can be equivalently decomposed to  \eqref{unsimp_admm1:eqn}-\eqref{unsimp_admm6:eqn}.
\end{proof}

\begin{proposition} \label{prop_simpequations}
Iterations \eqref{unsimp_admm1:eqn}-\eqref{unsimp_admm6:eqn} can be simplified to five steps resulting in:
\begin{equation}\label{simp_admm1:eqn}
\begin{aligned}
\Pi_{x}^t(k+1) \in \arg \min_{\Pi_{x}^t \in \mathcal{F}_{x}^t} -\sum_{y \in \mathcal{Y}_x} \delta_{xy} \pi_{xy}^t +\sum_{y \in \mathcal{Y}_x} \alpha_{xy}^t (k)\pi_{xy}^t \\ +\frac{\eta}{2} \sum_{y \in \mathcal{Y}_x}\left( \pi_{xy}^t - \pi_{xy}(k) \right)^2,
\end{aligned}
\end{equation}
\begin{equation}\label{simp_admm2:eqn}
\begin{aligned}
    \Pi_{x}^t(k+1) \in \arg \min_{\Pi_{x}^t \in \mathcal{F}_{x}^t} -\sum_{y \in \mathcal{Y}_x} \left( \delta_{xy} + \xi_{xy}' \right) \pi_{xy}^t \\ +\sum_{y \in \mathcal{Y}_x} \alpha_{xy}^t (k)\pi_{xy}^t + \frac{\eta}{2} \sum_{y \in \mathcal{Y}_x}\left( \pi_{xy}^t - \pi_{xy}(k) \right)^2,
\end{aligned}
\end{equation}
where we use \eqref{simp_admm1:eqn} for $x\in\mathcal{X}_o$ and \eqref{simp_admm2:eqn} for $x\in\mathcal{X}_a$.
\begin{equation}\label{simp_admm3:eqn}
\begin{aligned}
\Pi_{y}^s(k+1) \in \arg \min_{\Pi_{y}^s \in \mathcal{F}_{y}^s} -\sum_{x \in \mathcal{X}_y} \gamma_{xy} \pi_{xy}^s +\sum_{x \in \mathcal{X}_y} \alpha_{xy}^s(k)\pi_{xy}^s \\ +\frac{\eta}{2} \sum_{x \in \mathcal{X}_y} \left( \pi_{xy}(k) - \pi_{xy}^s \right),
\end{aligned}
\end{equation}
\begin{equation}\label{simp_admm4:eqn}
\begin{aligned}
    \pi_{xy}(k+1) = \frac{1}{2} \left( \pi_{xy}^t(k+1) + \pi_{xy}^s(k+1)\right),
\end{aligned}
\end{equation}
\begin{equation}\label{simp_admm5:eqn}
\begin{aligned}
    \alpha_{xy}(k+1) = \alpha_{xy}(k) + \frac{\eta}{2} \left( \pi_{xy}^t(k+1)-\pi_{xy}^s(k+1)\right).
\end{aligned}
\end{equation}
\end{proposition}

\begin{proof}
As \eqref{unsimp_admm4:eqn} is strictly concave, we can solve it by first-order condition:
$
    \pi_{xy}(k+1) = \frac{1}{2\eta}(\alpha_{xy}^{t}(k) - \alpha_{xy}^{s}(k)) + \frac{1}{2}(\pi_{xy}^{t}(k+1) + \pi_{xy}^{s}(k+1)).
$
By substituting the above equation into \eqref{unsimp_admm5:eqn} and \eqref{unsimp_admm6:eqn} we get:
$
    \alpha_{xy}^{t}(k+1) = \frac{1}{2}(\alpha_{xy}^{t}(k) + \alpha_{xy}^{s}(k)) + \frac{\eta}{2}(\pi_{xy}^{t}(k+1) - \pi_{xy}^{s}(k+1)),
$
$
    \alpha_{xy}^{s}(k+1) = \frac{1}{2}(\alpha_{xy}^{t}(k) + \alpha_{xy}^{s}(k)) + \frac{\eta}{2}(\pi_{xy}^{t}(k+1) - \pi_{xy}^{s}(k+1)).
$
We can see that $\alpha_{xy}^{t} = \alpha_{xy}^{s}$ during each update. Hence, $\pi_{xy}(k+1)$ can be further simplified as $\pi_{xy}(k+1) = \frac{1}{2} (\pi_{xy}^{t}(k+1) + \pi_{xy}^{s}(k+1))$ shown in \eqref{simp_admm4:eqn}. In addition, we can achieve \eqref{unsimp_admm5:eqn} and \eqref{unsimp_admm6:eqn} from  $\alpha_{xy}^{t} = \alpha_{xy}^{s} = \alpha_{xy}$ in \eqref{simp_admm5:eqn}.
\end{proof}

\begin{theorem}\label{them:pi_converge}
The algorithm described in Proposition \ref{prop_simpequations} converges to an optimal solution.
\end{theorem}

\begin{proof}
As \eqref{simp_admm1:eqn}-\eqref{simp_admm5:eqn} are equivalent to \eqref{unsimp_admm1:eqn}-\eqref{unsimp_admm6:eqn}, so it is sufficient to show that \eqref{unsimp_admm1:eqn}-\eqref{unsimp_admm6:eqn} converge to the optimal solution. The convergence of \eqref{unsimp_admm1:eqn}-\eqref{unsimp_admm6:eqn} directly follows from the general arguments in \cite[Section 3.2]{boyd2011distributed}. Therefore,
the iterations \eqref{simp_admm1:eqn} - \eqref{simp_admm5:eqn} converge to the optimal solution of \eqref{mainform:eqn}.
\end{proof}

In the above proposed distributed algorithm, each node computes its transport strategy based on the local information, i.e., information of connected nodes rather than all the nodes. The nodes update their strategies iteratively by communicating with connected neighbors. This is different from the centralized computation where the central planner needs to know all nodes' information to design the transport plan and then broadcasts the decision to the nodes.

\subsection{Integrated Distributed Algorithm}
We combine the algorithms for the attacker and the participants into one distributed algorithm. The integrated algorithm follows the updates below.
\begin{equation}\label{attacker_dist:eqn}
\begin{aligned}
    \bm{\xi}_x(k+1) \in & \arg\min_{\bm{\xi}_x,\bm{\chi}_x}\  \sum_{y\in\mathcal{Y}_x} \xi_{xy}\pi_{xy}(k)+ \bm{1}^{\mathsf{T}} \bm{\chi}_x
    \\
    \mathrm{s.t.}\quad 
    & \bm{\xi}_{x}\in\mathcal{A}_x,\  c_a\bm{\xi}_{x}\leq \bm{\chi}_x,\  c_a\bm{\xi}_{x}\geq -\bm{\chi}_x.
\end{aligned}
\end{equation}
\begin{equation}\label{combined_admm1:eqn}
\begin{aligned}
\Pi_{x}^t(k+1) \in \arg \min_{\Pi_{x}^t \in \mathcal{F}_{x}^t} -\sum_{y \in \mathcal{Y}_x} \delta_{xy} \pi_{xy}^t +\sum_{y \in \mathcal{Y}_x} \alpha_{xy}(k)\pi_{xy}^t \\+ \frac{\eta}{2} \sum_{y \in \mathcal{Y}_x}\left( \pi_{xy}^t - \pi_{xy}(k) \right)^2,\ \mathrm{for}\ x\in\mathcal{X}_o,
\end{aligned}
\end{equation}
\begin{equation}\label{combined_admm2:eqn}
\begin{aligned}
    \Pi_{x}^t(k+1) \in \arg \min_{\Pi_{x}^t \in \mathcal{F}_{x}^t} -\sum_{y \in \mathcal{Y}_x} \left( \delta_{xy} + \xi_{xy}(k) \right) \pi_{xy}^t \\ +\sum_{y \in \mathcal{Y}_x} \alpha_{xy}(k)\pi_{xy}^t + \frac{\eta}{2} \sum_{y \in \mathcal{Y}_x}\left( \pi_{xy}^t - \pi_{xy}(k) \right)^2,\ \mathrm{for}\ x\in\mathcal{X}_a,
\end{aligned}
\end{equation}
\begin{equation}\label{combined_admm3:eqn}
\begin{aligned}
\Pi_{y}^s(k+1) \in \arg \min_{\Pi_{y}^s \in \mathcal{F}_{y}^s} -\sum_{x \in \mathcal{X}_y} \gamma_{xy} \pi_{xy}^s +\sum_{x \in \mathcal{X}_y} \alpha_{xy}(k)\pi_{xy}^s \\ +\frac{\eta}{2} \sum_{x \in \mathcal{X}_y} \left( \pi_{xy}(k) - \pi_{xy}^s \right),
\end{aligned}
\end{equation}
\begin{equation}\label{combined_admm4:eqn}
\begin{aligned}
    \pi_{xy}(k+1) = \frac{1}{2} \left( \pi_{xy}^t(k+1) + \pi_{xy}^s(k+1)\right),
\end{aligned}
\end{equation}
\begin{equation}\label{combined_admm5:eqn}
\begin{aligned}
    \alpha_{xy}(k+1) = \alpha_{xy}(k) + \frac{\eta}{2} \left( \pi_{xy}^t(k+1)-\pi_{xy}^s(k+1)\right).
\end{aligned}
\end{equation}

The convergence of the integrated distributed algorithm is worth investigation. We have the following result.

\begin{theorem}
The designed integrated distributed algorithm \eqref{attacker_dist:eqn}-\eqref{combined_admm5:eqn} converges to a saddle-point equilibrium.
\end{theorem}

\begin{proof}
Based on Proposition \ref{prop:minimax_equi}, we know that there exists an equilibrium with $\{\bm{\xi}_x^*\}_{x\in\mathcal{X}_a}$ and $\Pi^*$ to the minimax game $G$. Theorem \ref{them:pi_converge} further shows that the max-problem \eqref{OTA_planner:eqn} converges to the best response of the min-problem \eqref{OTA_attacker_subp_2:eqn}.  Note that the trajectory of best response dynamics for continuous concave-convex zero-sum games always converges to saddle points \cite{hofbauer2006best}. Thus, the developed integrated distributed algorithm \eqref{attacker_dist:eqn}-\eqref{combined_admm5:eqn} converges to $\{\bm{\xi}_x^*\}_{x\in\mathcal{X}_a}$ and $\Pi^*$.
\end{proof}

For convenience, we summarize the integrated distributed algorithm in Algorithm \ref{Alg:1}.

\begin{algorithm}[!t]
\caption{Integrated Distributed Algorithm}\label{Alg:1}
\begin{algorithmic}[1]
\While {$\bm{\xi}_x$, $\Pi_{x}^t$ and $\Pi_{y}^s$ not converging}
\State Compute $\bm{\xi}_x(k+1)$ using \eqref{attacker_dist:eqn}, $\forall x \in \mathcal{X}_a$
\State Compute $\Pi_{x}^t(k+1)$  using \eqref{combined_admm1:eqn},  $\forall x\in \mathcal{X}_o$ 
\State Compute $\Pi_{x}^t(k+1)$  using \eqref{combined_admm2:eqn},  $\forall x\in \mathcal{X}_a$ 
\State Compute $\Pi_{y}^s(k+1)$  using \eqref{combined_admm3:eqn}, $\forall y\in\mathcal{Y}$
\State Compute $\pi_{xy}(k+1)$  using \eqref{combined_admm4:eqn}, $\forall \{x,y\}\in \mathcal{E}$
\State Compute $\alpha_{xy}(k+1)$  using \eqref{combined_admm5:eqn}, $\forall \{x,y\}\in \mathcal{E}$
\EndWhile
\State \textbf{return} $\bm{\xi}_x(k+1)$, $\forall x\in\mathcal{X}_a$ and $\pi_{xy}(k+1)$, $\forall \{x,y\}\in \mathcal{E}$
\end{algorithmic}
\end{algorithm}

\section{Case Studies}\label{sec:case}
\begin{figure}[!t]
    \centering
    \includegraphics[width=0.45 \columnwidth]{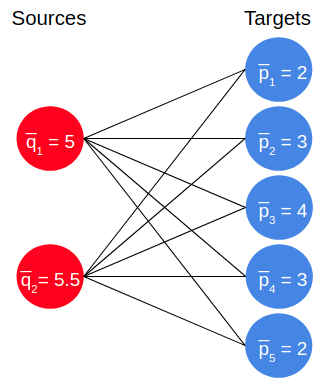}
    \caption{Bipartite transport network shows which source and target nodes are connected to one another.}
    \label{fig:f1}
\end{figure}

In this section we corroborate our algorithm for distributed OT while considering adversarial opponents. We consider the first case with five target nodes and two source nodes with a network structure connecting every source node to every target node as shown in Fig. \ref{fig:f1}. The upper bounds for the source nodes are $\bar{p}_1 = 2$, $\bar{p}_2 = 3$, $\bar{p}_3 = 4$, $\bar{p}_4 = 3$, $\bar{p}_5 = 2$, $\bar{q}_1 = 5$, and $\bar{q}_2 = 5.5$. The lower bound for all nodes are set to 0. Additionally, we consider linear utility functions $t_{xy}(\pi_{xy}) = \delta_{xy}\pi_{xy}$, and $s_{xy}(\pi_{xy}) = \gamma_{xy}\pi_{xy}, \forall\{x,y\} \in \mathcal{E}$. The corresponding parameters in the linear functions are selected as follows:
$$
    [\delta_{xy}]_{x\in\mathcal{X},y\in\mathcal{Y}} = \begin{bmatrix} 4 & 12 & 4 & 12 & 8 \\ 8 & 8 & 16 & 4 & 4\end{bmatrix},
$$ 
$$
    [\gamma_{xy}]_{x\in\mathcal{X},y\in\mathcal{Y}} = \begin{bmatrix} 6 & 4.5 & 12 & 6 & 9 \\ 3 & 6 & 7.5 & 9 & 12\end{bmatrix}.
$$
Furthermore, adversary's parameters are $c_a = 0.5$ and $\kappa_x = 15$, $\forall x\in\mathcal{X}_a$, and the deceptive targets include nodes 2 and 5. We next design the resilient transport strategy using the proposed distributed Algorithm \ref{Alg:1}.

\begin{figure}[!t] 
\centering
\subfigure[Social Utility]{\includegraphics[width=0.49\columnwidth]{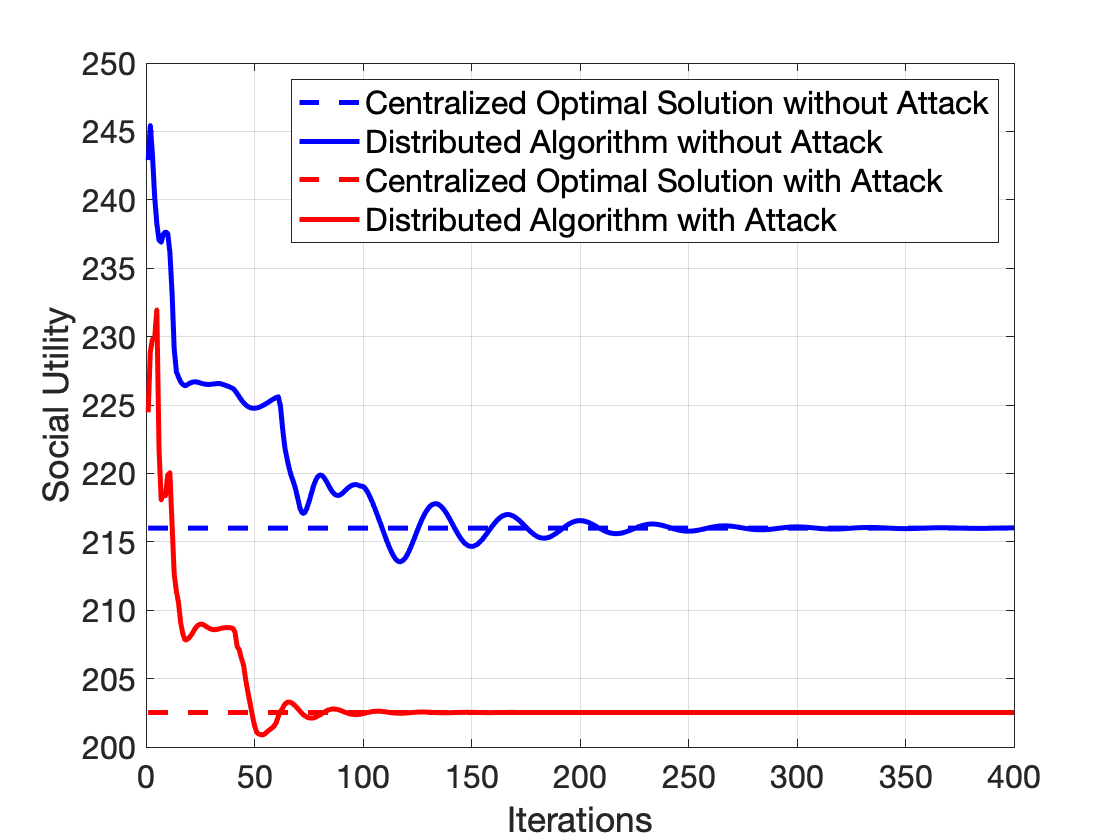}\label{fig:f2_1}}
\subfigure[Distance Residual]{\includegraphics[width=0.49\columnwidth]{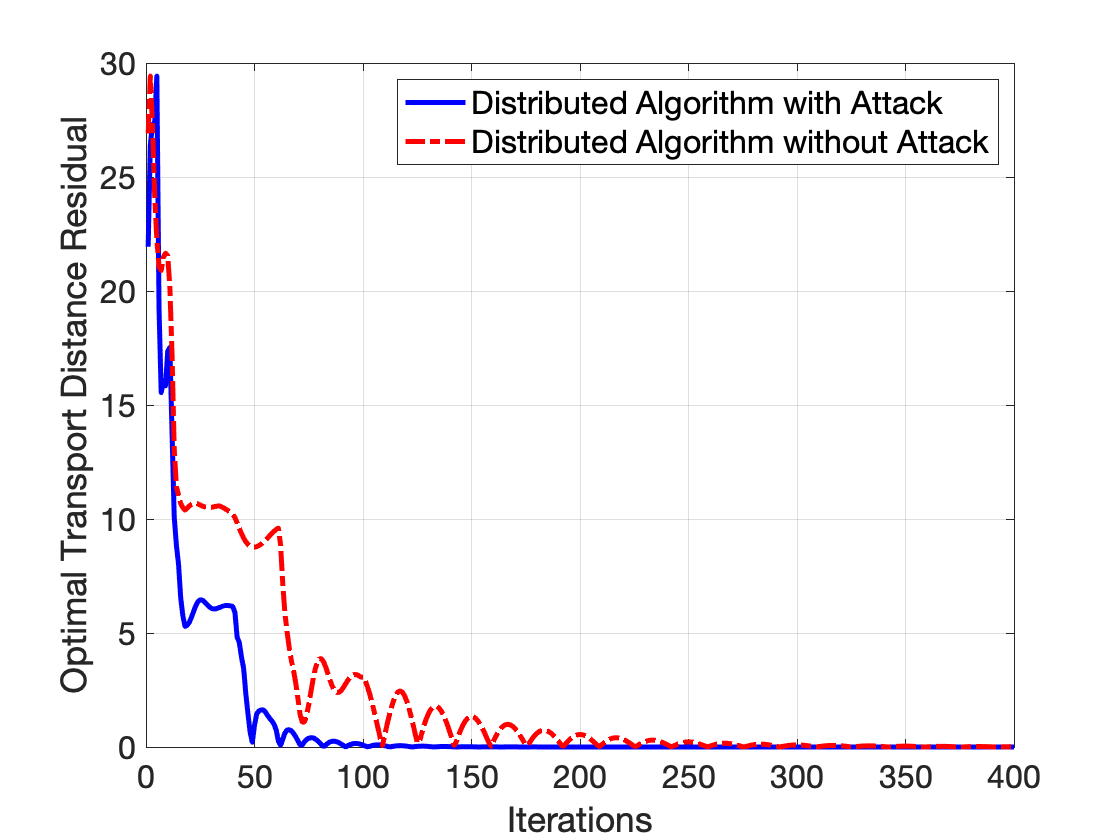}\label{fig:f2_2}}
\caption{Impact of the adversarial attacks on the transport strategy design using Algorithm \ref{Alg:1}. (a) and (b) depict the trajectories of social utility and residual of transport strategy, respectively.}
\label{fig:f2}
\end{figure}

\begin{figure}[!t] 
\centering
\subfigure[Attacker's Strategy]{\includegraphics[width=0.49\columnwidth]{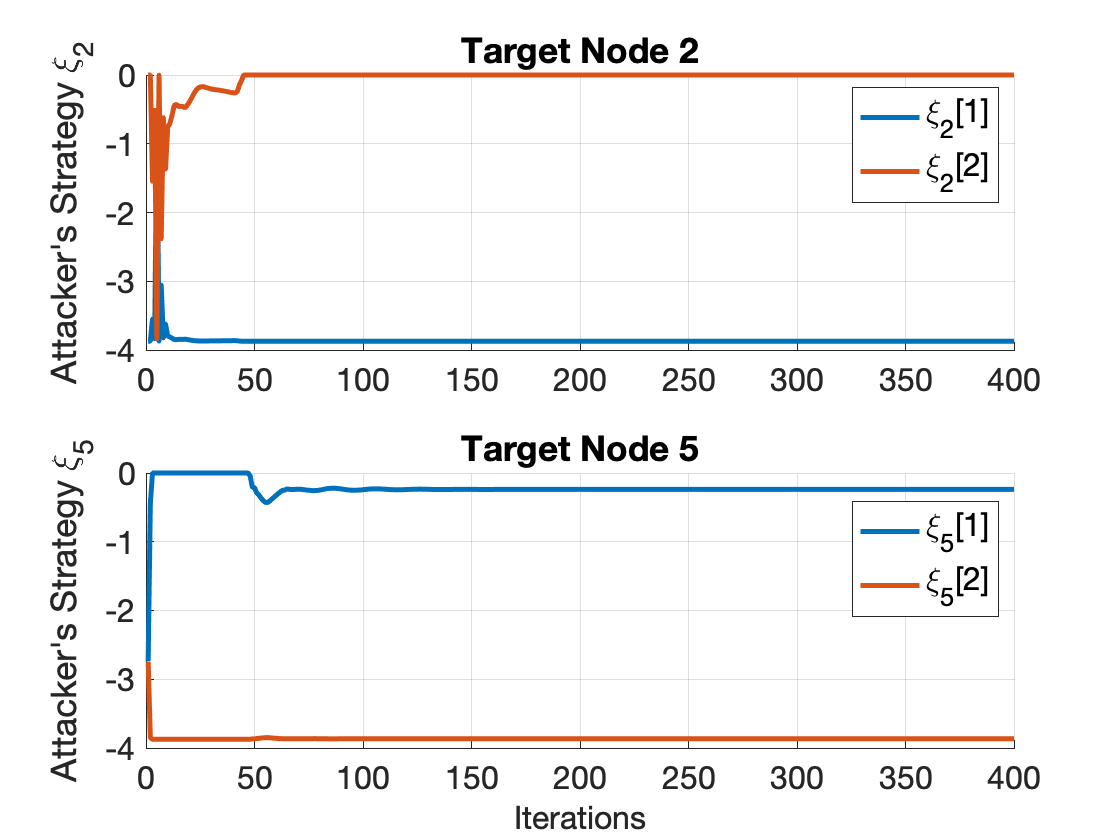}\label{fig:f3_1}}
\subfigure[Transport Plan]{\includegraphics[width=0.49\columnwidth]{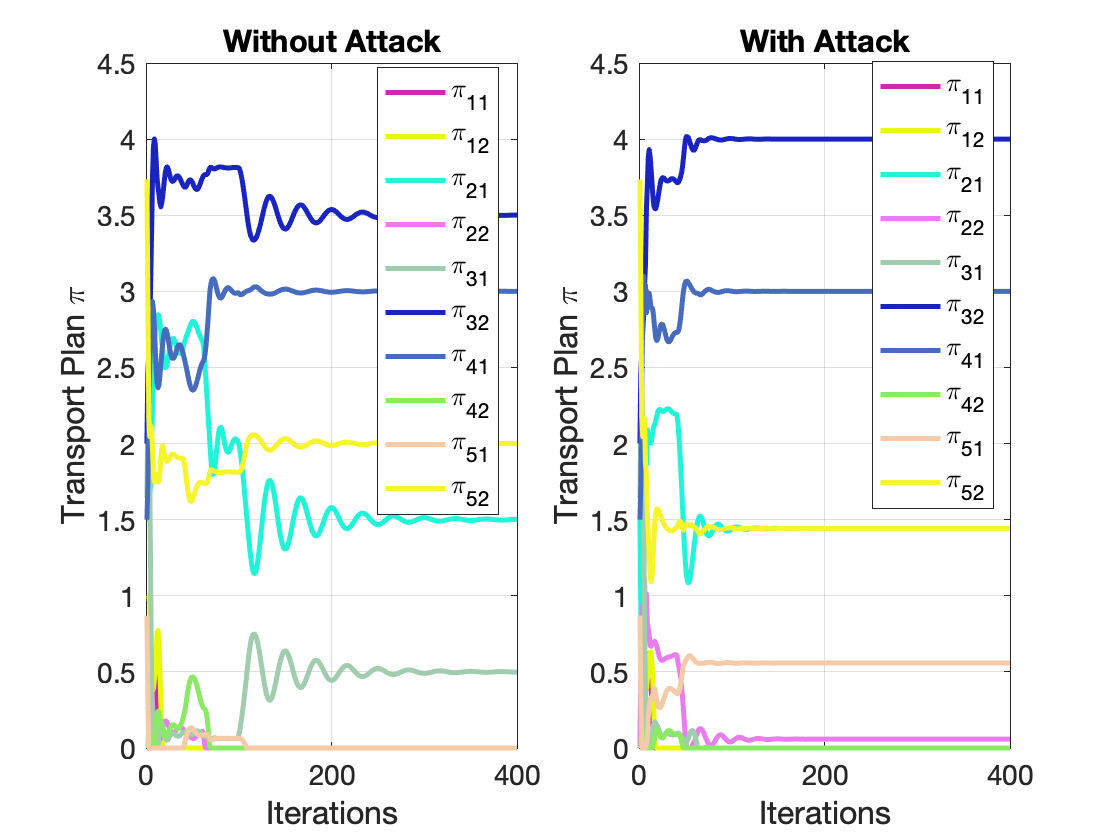}\label{fig:f3_2}}
\caption{(a) shows the attacker's strategy at the target nodes 2 and 5. (b) shows the corresponding transport plan under two scenarios.}
\label{fig:f3}
\end{figure}

First, we show that the algorithm works and converges to the same value obtained by the centralized method. We also compare the transport strategies when the network with and without adversaries. When there is an adversary, we use a combination of \eqref{combined_admm1:eqn} (for benign targets) and \eqref{combined_admm2:eqn} (for deceptive targets) to calculate $\Pi_x^t(k+1)$. When there is no adversary, meaning none of the nodes are compromised, we only use \eqref{combined_admm2:eqn} to compute $\Pi_x^t$.
The results are shown Fig. \ref{fig:f2}. Specifically, Fig. \ref{fig:f2_1} shows the social utility which is the aggregated payoff all nodes. Fig. \ref{fig:f2_1} corroborates that the algorithm converges to the centralized solution in both scenarios with and without attacks. We also note that when we consider an attack the algorithm converges to a lower social utility. This is due to the fact that we have to account for the adversarial impacts which decreases the desired utility between the source node and the compromised target node. \ref{fig:f2_2} highlights the distance residual of the transport strategy, which measures the difference between the strategy at each step and the equilibrium solution. 
The attacker's strategy $\bm{\xi}_x$ is shown in Fig. \ref{fig:f3_1}. For both compromised nodes, the deceptive strategies $\bm{\xi}_2$ and $\bm{\xi}_5$ converge to a nonzero values, indicating that the attacker is actively affecting the transport plan. Fig. \ref{fig:f3_2} further illustrates this phenomenon as the resource allocation strategies are different in the two investigated cases. 

\begin{figure}[!t] 
\centering
\subfigure[Social Utility]{\includegraphics[width=0.49\columnwidth]{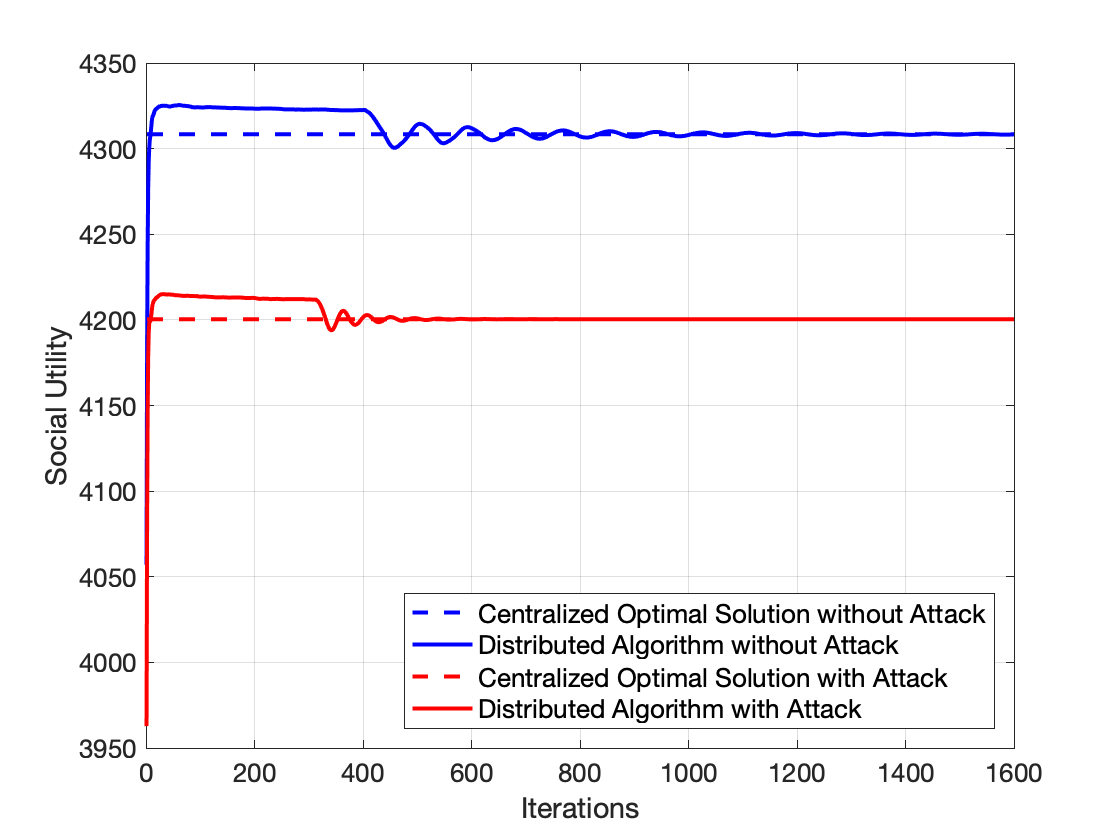}\label{fig:case2_1}}
\subfigure[Distance Residual]{\includegraphics[width=0.49\columnwidth]{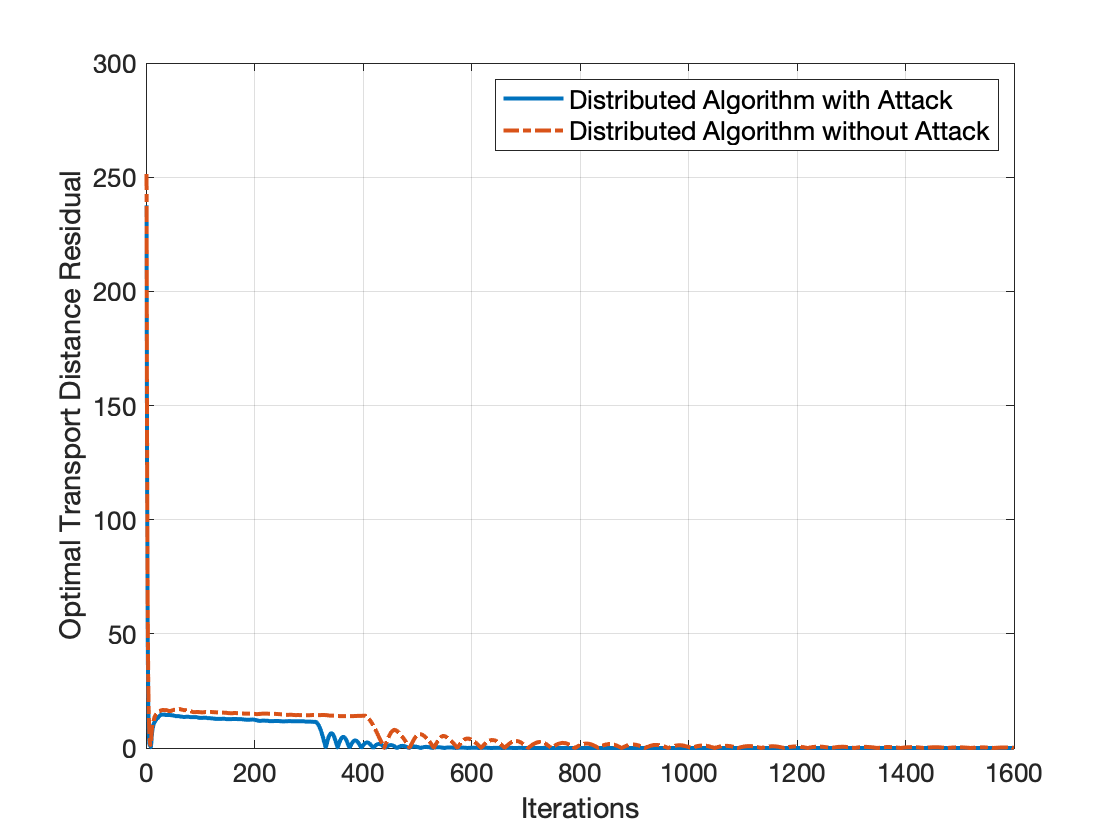}\label{fig:case2_2}}
\subfigure[Attacker's Strategy]{\includegraphics[width=0.49\columnwidth]{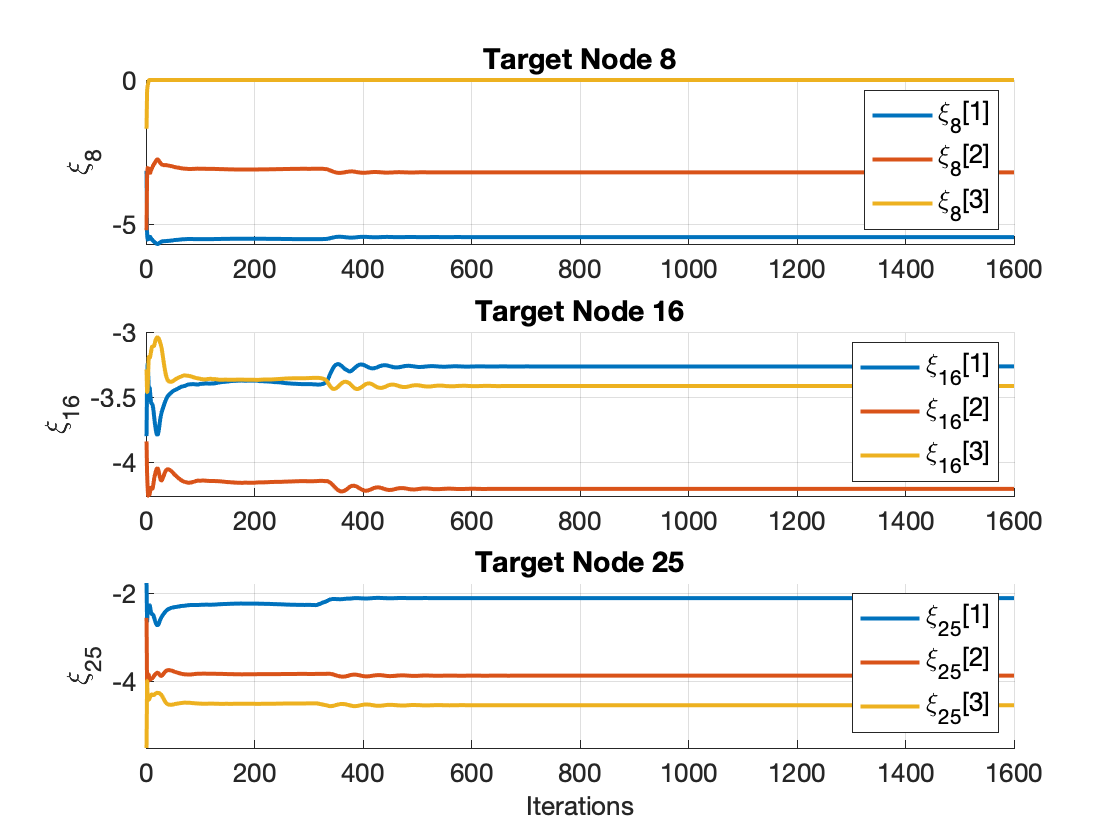}\label{fig:case2_3}}
\subfigure[Transport Plan]{\includegraphics[width=0.49\columnwidth]{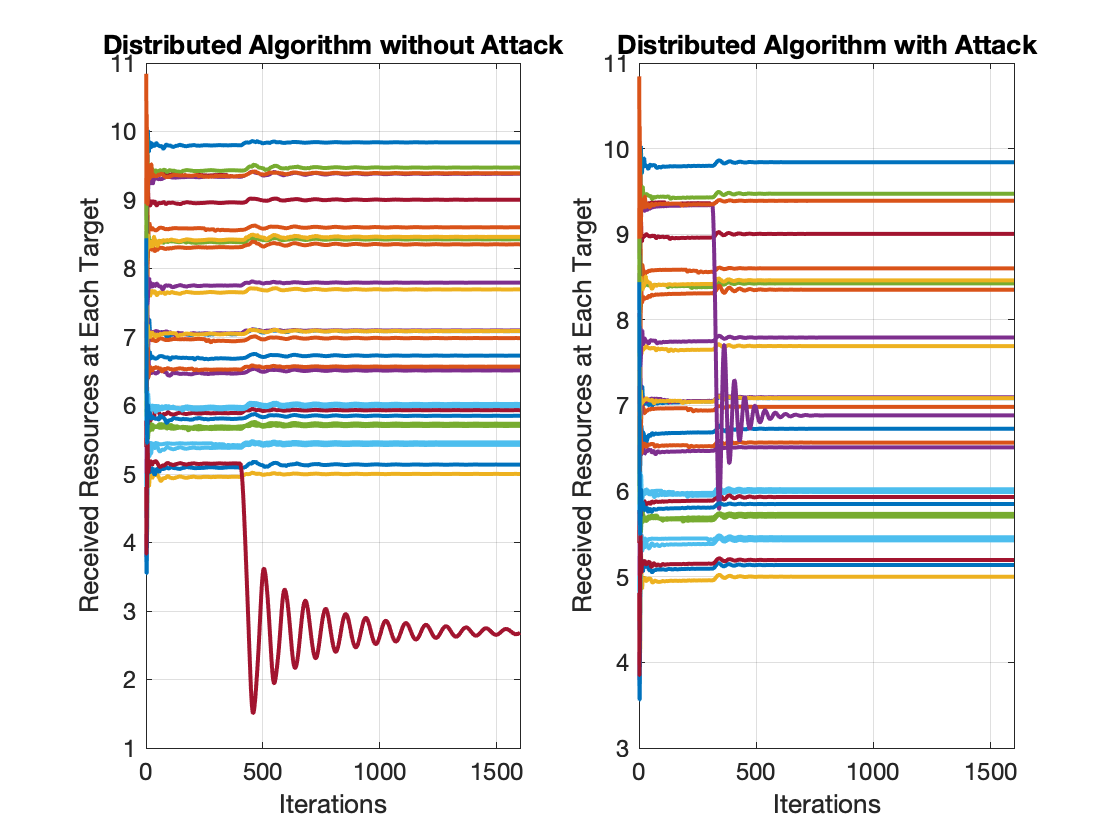}\label{fig:case2_4}}
\caption{Example of a larger-scale network. (a) and (b) depict the trajectories of social utility and residual of transport strategy, respectively. (c) and (d) show the attacker's strategy and the corresponding transport plans, respectively.}
\label{fig:case2}
\end{figure}


We further investigate a larger scale network with 3 source nodes and 30 target nodes and every target is connected to every source node. The parameters are generated randomly following uniform distributions: $\delta_{xy}\sim U(6,11)$, $\gamma_{xy}\sim U(7,12)$, $\bar{p}_x\sim U(5,10)$, and $\bar{q}_y\sim U(67,75)$. Nodes 8, 15, and 25 are considered to be possibly compromised with $c_a=0.5$ and $\kappa_x=40$. The obtained results are shown in Fig. \ref{fig:case2}. The results also converge to the centralized solutions. We can conclude that the designed algorithm is applicable to large-scale networks.

\section{Conclusion}\label{sec:conclusion}
In this paper, we have investigated an adversarial discrete optimal transport framework for resource matching in which the participating nodes could be malicious by reporting untruthful preference parameters. We have developed a distributed algorithm for computing the strategic resource allocation strategies which are resilient to such attacks. The designed algorithm converges to a same solution as one designed by a centralized planner, and it is applicable to large scale networks susceptible to deceptive attacks. The adversarial behavior is specifically acknowledged in the algorithm when a participating node is compromised. Each connected pair of target and source nodes negotiate on the their proposed transport plans, and thus the compromised node's actions is taken into account in the final allocation schemes. 
The algorithm terminates when the sources and targets reach a consensus.
Future work includes to consider the differential privacy of the nodes in the network when designing the algorithm. Another direction is to develop a formal metric to quantify the stealthiness of the attacker and integrate it with the established adversarial optimal transport framework.

\bibliographystyle{IEEEtran}
\bibliography{references.bib}

\begin{thebibliography}{10}
\providecommand{\url}[1]{#1}
\csname url@samestyle\endcsname
\providecommand{\newblock}{\relax}
\providecommand{\bibinfo}[2]{#2}
\providecommand{\BIBentrySTDinterwordspacing}{\spaceskip=0pt\relax}
\providecommand{\BIBentryALTinterwordstretchfactor}{4}
\providecommand{\BIBentryALTinterwordspacing}{\spaceskip=\fontdimen2\font plus
\BIBentryALTinterwordstretchfactor\fontdimen3\font minus
  \fontdimen4\font\relax}
\providecommand{\BIBforeignlanguage}[2]{{%
\expandafter\ifx\csname l@#1\endcsname\relax
\typeout{** WARNING: IEEEtran.bst: No hyphenation pattern has been}%
\typeout{** loaded for the language `#1'. Using the pattern for}%
\typeout{** the default language instead.}%
\else
\language=\csname l@#1\endcsname
\fi
#2}}
\providecommand{\BIBdecl}{\relax}
\BIBdecl

\bibitem{galichon2016econ}
A.~Galichon, \emph{Optimal Transport Methods in Economics}.\hskip 1em plus
  0.5em minus 0.4em\relax Princeton University Press, 2016.

\bibitem{bayat2016matching}
S.~Bayat, Y.~Li, L.~Song, and Z.~Han, ``Matching theory: Applications in
  wireless communications,'' \emph{IEEE Signal Processing Magazine}, vol.~33,
  no.~6, pp. 103--122, 2016.

\bibitem{zhang2019consensus}
R.~Zhang and Q.~Zhu, ``Consensus-based distributed discrete optimal transport
  for decentralized resource matching,'' \emph{IEEE Transactions on Signal and
  Information Processing over Networks}, vol.~5, no.~3, pp. 511--524, 2019.

\bibitem{jhughes2021fair}
J.~Hughes and J.~Chen, ``Fair and distributed dynamic optimal transport for
  resource allocation over networks,'' in \emph{55th Annual Conference on
  Information Sciences and Systems (CISS)}, 2021.

\bibitem{basar1998dynamic}
T.~Ba{\c{s}}ar and G.~J. Olsder, \emph{Dynamic Noncooperative Game
  Theory}.\hskip 1em plus 0.5em minus 0.4em\relax SIAM, 1998.

\bibitem{garnaev2014fair}
A.~Garnaev and W.~Trappe, ``Fair resource allocation under an unknown jamming
  attack: a bayesian game,'' in \emph{IEEE International Workshop on
  Information Forensics and Security (WIFS)}, 2014, pp. 227--232.

\bibitem{shao2020distributed}
G.~Shao, R.~Wang, X.-F. Wang, and K.-Z. Liu, ``Distributed algorithm for
  resource allocation problems under persistent attacks,'' \emph{Journal of the
  Franklin Institute}, vol. 357, no.~10, pp. 6241--6256, 2020.

\bibitem{chen2016joint}
H.~Chen, M.~Zhou, L.~Xie, K.~Wang, and J.~Li, ``Joint spectrum sensing and
  resource allocation scheme in cognitive radio networks with spectrum sensing
  data falsification attack,'' \emph{IEEE Transactions on Vehicular
  Technology}, vol.~65, no.~11, pp. 9181--9191, 2016.

\bibitem{boyd2011distributed}
S.~Boyd, N.~Parikh, and E.~Chu, \emph{Distributed Optimization and Statistical
  Learning via the Alternating Direction Method of Multipliers}.\hskip 1em plus
  0.5em minus 0.4em\relax Now Publishers, 2011.

\bibitem{nikaido1954neumann}
H.~Nikaid{\^o}, ``On von {N}eumann's minimax theorem,'' \emph{Pacific Journal
  of Mathematics}, vol.~4, no.~1, pp. 65--72, 1954.

\bibitem{hofbauer2006best}
J.~Hofbauer and S.~Sorin, ``Best response dynamics for continuous zero--sum
  games,'' \emph{Discrete \& Continuous Dynamical Systems-B}, vol.~6, no.~1, p.
  215, 2006.

\end{thebibliography}

\end{document}